\documentclass{emulateapj-rtx4}

\shorttitle{bursts reveal a tiny life cycle of corona}
\shortauthors{Chen et al.}

\begin{document}

\title{Type-I X-ray bursts reveal a fast co-evolving behavior of  the corona in an X-ray binary}

\author{
Yu-Peng Chen\altaffilmark{1}, Shu Zhang\altaffilmark{1}, Shuang-Nan Zhang\altaffilmark{1},
Jian Li\altaffilmark{1},
Jian-Min Wang\altaffilmark{1,2}
}

\altaffiltext{1}
{
Key Laboratory for Particle Astrophysics, Institute of High
Energy Physics, Chinese Academy of Sciences, 19B Yuquan Road,
Beijing 100049, China; chenyp@ihep.ac.cn,
szhang@ihep.ac.cn, zhangsn@ihep.ac.cn
}

\altaffiltext{2}
{
Theoretical Physics Center for Science Facilities (TPCSF), CAS
}

\begin{abstract}
The coronae in X-ray binaries (XRBs) still remain poorly understood, although they have been believed for a long
time to play a key role in modeling the characteristic outbursts of XRBs. Type-I X-ray bursts, the
thermonuclear flashes happening on the surface of a neutron star (NS), can be used as a probe to the innermost
region of a NS XRB, where the corona is believed to be located very close to the NS. We report the discovery of
a tiny life cycle of the corona that is promptly co-evolved with  the type-I bursts superimposed on the outburst of the NS XRB
IGR~J17473$-$2721. This finding may serve as the first evidence of directly seeing the rapid disappearance and
formation of a corona in an XRB with a cooling/heating timescale of less than a second, which can strongly
constrain the accretion models in XRBs at work.
\end{abstract}

\section{Introduction}

The spectra of
low-mass X-ray binaries (LMXBs), including both black hole (BH) and neutron star (NS) XRBs, are generally
disentangled into soft/thermal (e.g., a blackbody-like) and hard/Comptonized (e.g., power-law shape with a
cutoff at tens to hundreds keV) components. For example, a hard X-ray power law with a cutoff energy of tens keV
in a NS LMXB was first observed in the atoll source 4U~1608$-$522 \citep{zhang1996}. A corona in a region close
to the compact object of an XRB is usually invoked to account for the hard X-rays produced via inverse Compton
scattering of the soft photons off the hot electrons in the corona. Although the existence for such a corona has been
hinted from time to time, so far a direct observational evidence of its existence is still missing. It also
remains a puzzle on how the corona is formed and where it is located. Plasmas may be energized to form a corona
via either evaporation  \citep{meyer1994,Esin1997,liu2007,frank2002} or magnetic re-connections  \citep{zhang2000,zhang2007,mayer2007}. The corona's location is still not known
for NS XRBs, and accordingly the Eastern \citep{mitsuda89} and the Western models \citep{white88} were proposed.
The Eastern model considers a corona around the NS, whereas the Western model takes the corona covering the
accretion disk.
A hybrid model \citep{lin07} for NS X-ray binaries \textbf{was} devised  by analyzing the X-ray spectra of Aql X-1 and 4U 1608-52, and  offered  a weak-Comptonization solution that differs from the above two models.
X-ray binaries (XRBs) are commonly observed with outbursts displaying strong spectral evolution.
Here we take advantage of the type-I X-ray bursts from the hard surface of a NS to probe the
purported corona, under a circumstance of outburst and  accompanying spectral evolution. Type-I X-ray bursts are caused by unstable burning of the accreted hydrogen/helium on the
surface of a NS, and manifest themselves as a sudden increase (typically by a factor of 10 or greater) in the
X-ray luminosity (for reviews, see \citep{Lewin,Cumming,Strohmayer,Galloway}).
\section{Observations and data analysis}

IGR~J17473$-$2721 was discovered by INTEGRAL during an outburst in April 2005 \citep{Grebenev}. Its subsequent
outburst in 2008 March - October was almost uniformly monitored by RXTE/PCA (Proportional Counters Array),
resulting in 182 RXTE/PCA pointed observations, with the identifier (OBSID) of proposal number (PN) 93064,
93093, and 93442 in the
 High Energy Astrophysics Science Archive Research Center (HEASARC).
These observations adopted in this analysis cover the entire outburst and add up to $\sim$ 475 ks of exposure
time on the source. The analysis of the PCA data is performed by using
 HEAsoft v. 6.6. The data are filtered using the standard RXTE/PCA criteria.
 Only the data from the PCU2 (the third Proportional Counters Unit, in the 0-4 numbering scheme) are used for the analysis,
 because this PCU was 100\% working during all the
 observations. The background file used in the analysis of PCA data is the most recent one for
bright-sources found at the HEASARC website\footnote{pca$\_$bkgd$\_$cmbrightvle$\_$eMv20051128.mdl}. The dead
time correction is made to all the spectra and lightcurves under the standard procedure described at the HEASARC
website\footnote{http:$//$heasarc.nasa.gov$/$docs$/$xte$/$recipes$/$pca$\_$deadtime.html}.

\section{Results}

\begin{figure}[t]
\centering
  \includegraphics[angle=0, scale=0.8] {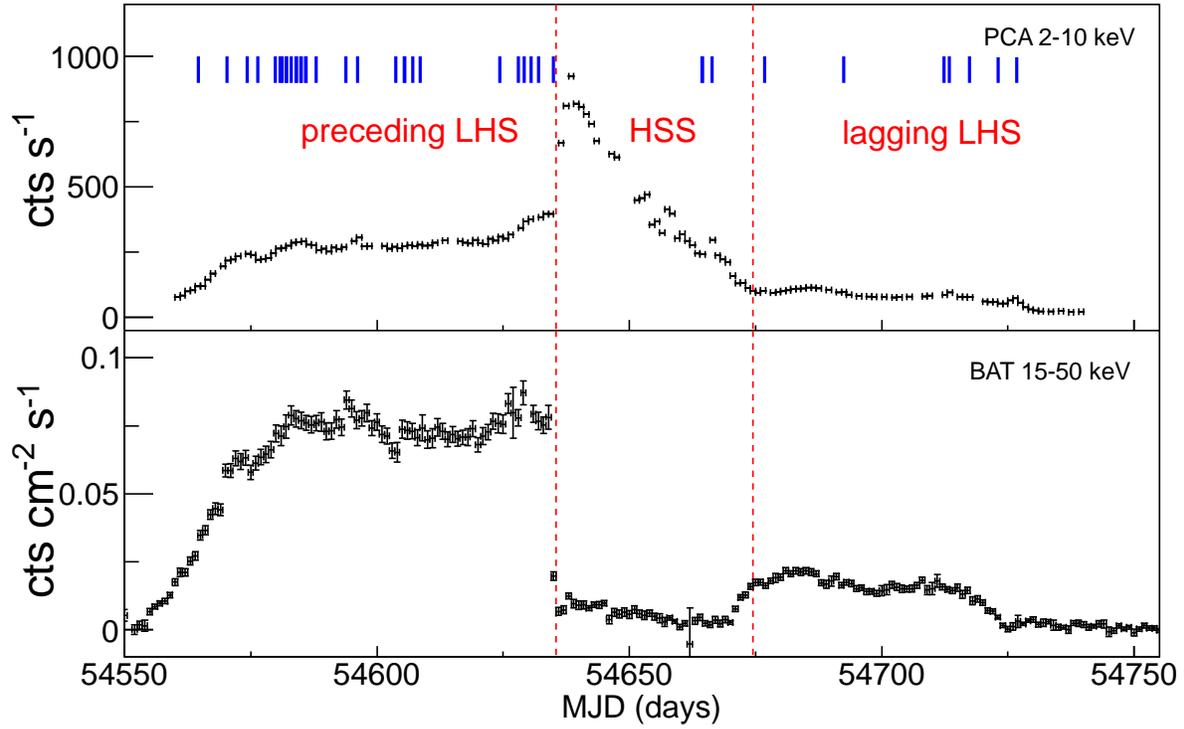}
 \caption{RXTE/PCA lightcurve (2-10 keV, upper
panel) and Swift/BAT lightcurve (15-50 keV, lower panel) covering the 2008 outburst of IGR~J17473$-$2721 with a
time resolution of 1 day. The dashed lines show the different states of the outburst. The locations of the
bursts are marked at the top of the upper panel (blue vertical lines).}
 \label{lc_outburst}
\end{figure}

As shown in Figure \ref{lc_outburst} the source in the 2008 outburst experienced a two-months preceding low/hard
state (LHS) and a lagging LHS with respect to the high/soft state (HSS); here `low' and `high' refers to its
flux in the 2-10 keV band, and `soft' and `hard' refers to its spectral shape from 2-50 keV.
The preceding LHS and  the lagging LHS are usually refereed in literature as  the high luminosity LHS and  the low luminosity LHS, respectively \citep{Remillard2006}.
The flux of the
lagging LHS is roughly 4 times lower than the preceding LHS, presenting with a so-called hysteresis typical to the
outbursts of BH XRBs \citep{miyamoto95,maccarone2003,meyer2005,dunn2010,tang2011}. Besides
4U~1908+005 (Aql~X$-$1) and 4U~1608$-$522 \citep{Gladstone2007}, IGR~J17473$-$2721 is thus the third NS XRB
observed with hysteresis. There are 32 bursts located in the preceding LHS and 10 bursts in the tail of the HSS
and the lagging LHS \citep{chen2010,chen2011,Chenevez2011}, constituting an entire sample of 42 bursts recorded by RXTE. The long-lived preceding LHS and the large number of
burst events occurred in both the preceding and lagging LHSs make IGR~J17473$-$2721 almost unique for probing
its corona \citep{zhang2009} under a circumstance of an outburst evolution.

For both LHSs, the persistent emission can be fitted by a power-law model with a cutoff energy $\sim$40 keV. The
persistent luminosity in the preceding LHS and the lagging LHS are $\sim$0.1 ${L}_{\rm Edd}$ (Eddington
luminosity) and $\sim$0.025 ${L}_{\rm Edd}$, respectively. The burst spectrum is well modeled by a blackbody
model of a characteristic temperature of less than 3 keV. The burst emission can reach ${L}_{\rm Edd}$, and
dominates the total emission at energies well below $\sim$30 keV, above which the persistent emission from the
purported corona dominates. We therefore take the emission in the 30-50 keV energy band to investigate the
possible influence of the bursts upon the corona.

\begin{figure}[t]
\centering
  \includegraphics[angle=0, scale=0.35] {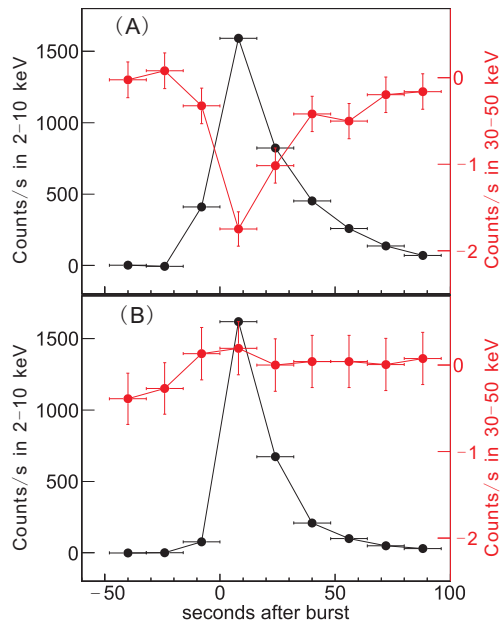}
  \caption{The 16s-bin lightcurves for the bursts in the preceding LHS (upper panel) and the lagging LHS (lower panel). Each data point is the sum over the spectral residual after subtracting off the persistent emissions at 2-10 keV (black) and 30-50 keV (red), respectively.}
\label{spe_lc_1}
\end{figure}

We choose 40 bursts out of the entire sample recorded by RXTE in the 2008 outburst with complete data for both
the burst and the persistent emission. For each burst, we use the time when the burst reached its peak at 2-10
keV as a reference to produce the lightcurve and spectrum of each burst. Those X-rays recorded 48 seconds before
and 80 seconds after the reference time are regarded as the background and are subtracted off for each burst in timing and
spectral analysis. The 30-50 keV persistent count rate recorded by RXTE/PCA in  the preceding
LHS, the lagging LHS and the HSS are $\sim$4.0 cts/s, $\sim$1.0 cts/s, and $\sim$0.11 cts/s, respectively. With
respect to the reference time we split each burst into a sequence of 16-second segments.
After the persistent emission is subtracted off, bursts are combined for those located in the preceding LHS or in the lagging LHS.
 As shown in Figure \ref{spe_lc_1}, the 30-50 keV flux of the combined burst in the
preceding LHS is mostly negative during the burst and around zero elsewhere; however, the fluxes of the combined
burst in the lagging LHS are always consistent with zero.

\begin{figure}[t]
\centering
   \includegraphics[origin=c, angle=0, scale=0.5] {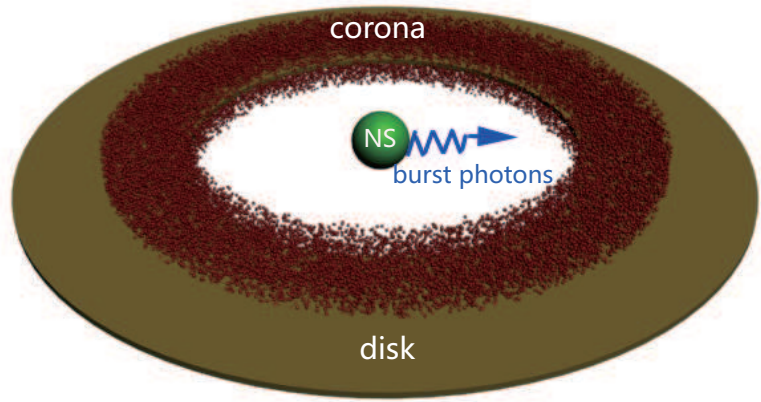}
 \caption{The illustration of the central region of a NS XRB, in which a corona is located around the disk and cooled by the soft X-rays from a type-I burst occurred on surface of the NS.}
\label{ill}
\end{figure}

\begin{figure}[t]
\centering
   \includegraphics[angle=0, scale=0.45] {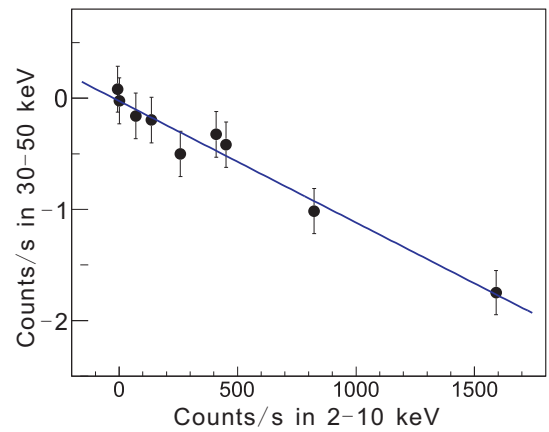}
 \caption{A linear fit on the data shown in the upper panel of Figure 2.}
\label{spe_lc_2}
\end{figure}

\begin{figure}[t]
\centering
   \includegraphics[angle=270, scale=0.3] {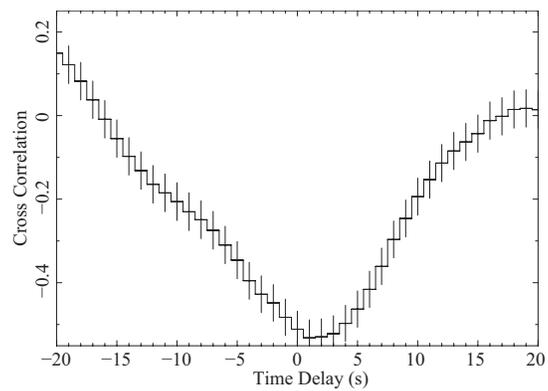}
 \caption{The cross-correlation between the 2-10 keV and 30-50 keV, with a time resolution of 1 second, for the combined burst in the preceding LHS.}
\label{spe_lc_3}
\end{figure}

Our previous reports showed that   the corona of the system (Figure \ref{ill}) is most likely located around a largely inclined disk and forms a relatively small opening angle with respect to the surface of neutron star,
so that the type-I X-ray bursts can escape from the NS surface without suffering severe Comptonization \citep{zhang2009,chen2011}.
The 30-50 keV decrement  reaches a maximum of $\sim$2
cts/s at the 2-10 keV burst peak, accounting to about half of the 30-50 keV persistent flux. This suggests that
half of the corona were cooled by the soft photons of the bursts. Actually the 30-50 keV profile in the preceding LHS
is anti-correlated with that of the 2-10 keV under a correlation coefficient of 0.89 (see Figure
\ref{spe_lc_2}). In this analysis, a time resolution of 16 seconds is used to show the significance of the flux decrement,
due to the very low flux in the 30-50 keV band. We try as well to extract the net lightcurves of the combined burst in the
2-10 keV and 30-50 keV bands with a time resolution of 1 second in the preceding LHS. A cross-correlation
analysis between the two lightcurves shows that the 30-50 keV X-rays lag the 2-10 keV X-rays by 0.7$\pm$0.5
second, as shown in Figure \ref{spe_lc_3}. This suggests that the fading and recovering of the corona follow the
burst flux change almost instantaneously. In the preceding LHS, the Compton cooling time of a pure
electron-positron plasma is $t_{\rm compt}=E_{\rm e}/P_{\rm compt}=5.4\times10^{-8}$~s, here $E_{\rm e}\sim 20
~{\rm keV}$ is the electron energy. However, the plasma is more likely a proton-electron plasma, whose cooling
time is $\sim10^{-7}$~s, i.e., increased by a factor of 2. This means that the heating time of the corona must
be longer than $\sim10^{-7}$~s; otherwise the corona cannot be cooled down as observed during the rising phase
of the bursts. On the other hand, the heating time must be shorter than about 1 second; otherwise the recovering
of the corona would lag the bursts in the falling phase. These results serve as the first observational evidences in an XRB for a promptly cooling and recovering corona which manifests itself with an apparent  tiny life cycle via tightly co-evolving with the soft X-ray bursts.

It is normally difficult to estimate the mass of the corona in an XRB, because the seed photon energy density
$U_{\rm ph}$ for the inverse Compton scattering in the corona cannot be observed directly and cleanly.
Fortunately, the prompt response of the corona to the bursts offers an unique opportunity to estimate the mass
of the corona directly. The temperature and Compton scattering optical depth of the corona are obtained to be
around 20 keV and 2 for both LHSs, respectively, by fitting their persistent emission spectra with the thermal
Comptonization model ({\it Comptt} in Xspec)\citep{titarchuk1994}. With the bursts as the coolant, the seed
photon energy density is give by,
\begin{equation}
U_{\rm ph}=\frac{L_{\rm burst}}{4\pi cR^{\rm 2}},
\end{equation}
here $R$ is the distance of the corona from the NS. The total power for a single scattering from a
non-relativistic electron with temperature $kT$, i.e., the cooling rate, is given by,
\begin{equation}
P_{\rm compt}=(\frac{4kT}{mc^{\rm 2}})c\sigma_{\rm T}U_{\rm ph}\ ({\rm erg~s^{-1}}),
\end{equation}
here $\sigma_{\rm T}$ and $U_{\rm ph}$ are the Thomson cross section and photon energy density of a burst in the
corona, respectively. The mass of the corona cooled by a burst can be calculated as,
\begin{equation}
\Delta M_{\rm corona}=\frac{\Delta L_{\rm LHS}}{P_{\rm comp}}m_{\rm H}.
\end{equation}
The inner radius of the accretion disk is assumed as the radius of the magnetosphere of the NS, where the gas
pressure in the disk is balanced by the magnetic pressure of magnetosphere \citep{cui1997,weng2011}. Taking
$L_{\rm burst}=8.8\times10^{37}$~erg s$^{-1}\sim 0.5~L_{\rm Edd}$ , $L_{\rm LHS}= 1.8\times10^{37}$~erg
s$^{-1}\sim 0.1~L_{\rm Edd}$, the mass of NS $\sim 1.4~M_{\bigodot}$, the NS surface magnetic field strength
$\sim 10^{8}~G$, the radius of the NS $\sim 10$~km, we have, $R=12~{\rm km}$ and $M_{\rm corona}=2\Delta M_{\rm
corona}=5.8\times10^{13}~{\rm g}$. Since the Compton scattering optical depth is around 2 and in the above
calculation a single scattering is assumed, the calculated corona mass is only an upper limit. The accretion
rate in the preceding LHS $\dot{M}=\frac{L_{\rm LHS}}{\epsilon c^{\rm 2}}=1.0\times10^{17}~{\rm g}~{\rm
s}^{-1}$, here $\epsilon=0.2$ is the radiative efficiency for a NS XRB. Therefore the accretion process has
sufficient mass reservoir to supply the required mass to the corona almost instantaneously, as observed.

\section{Discussion}

Despite the similar spectral properties of the persistent emission for both the preceding and lagging LHSs, no
decrement of the 30-50 keV photons is observed in the combined burst in the lagging LHS. The lower flux in the
lagging LHS indicates an accretion rate decreasing by a factor of 4 compared to that in the preceding LHS. Therefore
the inner disk radius ($R\propto L^{-2/7}$)\citep{weng2011} in the LHS is almost doubled, resulting a lower cooling rate by the
bursts. With the persistent flux of about 1 cts/s in the lagging LHS in the 30-50 keV
band, the maximum of decrement in the combined burst is expected to be $\sim$0.2~cts/s, well within the observed
uncertainties. This explains the lack of flux decrement in the 30-50 keV band in the lagging LHS.

Spectral transitions in XRB outbursts have been studied in the frameworks of producing/supressing the hard
X-rays via mechanisms, e.g., disk viscous dissipation and Compton cooling, where one key parameter is the time
scale of flux variability. While the inverse Compton scattering has been applied successfully to account for the
cooling, however no consensus has been reached on the correct corona formation model, e.g., between disk
evaporation and magnetic re-connection models. Unfortunately, the time scales for either the transition into a
HSS or stepping back to a LHS have so far only been observed not shorter than days, not sufficiently fine to
discriminate between different models. For the disk evaporation, the formation of a corona is driven and
energized by the disk accretion, which has a typical time scale of days
\citep{meyer1994,Esin1997,liu2007,frank2002}, obviously inconsistent with the observed heating time of shorter than
1 second. On the other hand, magnetic field reconnections in the inner disk region can release the kinetic
energies in the rotating disk to heat the corona with the Keplerian orbital time scale of the order of
milliseconds.

Therefore microscopic processes in the accretion disk should be responsible for forming the coronae in this NS
XRB, in a process similar to that powering the solar flares in the Sun, as already discussed previously for BH
accreting systems \citep{zhang2000,mayer2007,zhang2007}. Future X-ray instruments with fine timing resolution and
much larger effective areas than the RXTE/PCA instrument, may allow direct probing  the fine details of the
microscopic processes in the accretion and corona formation, by observing the interplays between type I X-ray
bursts and the rapid spectral evolution of NS XRBs.

\acknowledgements
This work is supported in part by the National Natural Science Foundation of China, the CAS key Project
KJCX2-YW-T03, 973 program 2009CB824800 and NSFC-11103020, 11133002, 10725313, 10521001, 10733010, 10821061,
11073021, 11173023. This research has made use of data obtained from the High Energy Astrophysics Science
Archive Research Center (HEASARC), provided by NASA's Goddard Space Flight Center.


\clearpage

\end{document}